\documentstyle[12pt]{article}
\begin{document}
\begin{flushright}
WU-HEP-95-1 \\
TWC-95-1 \\
January 1995
\end{flushright}
\begin{center}
\begin{Large}
Numerical Analyses of CERN 200GeV/A Heavy-Ion Collisions

\vspace{.2cm}
Based on a Hydrodynamical Model with Phase Transition
\\
\end{Large}
\vspace{1cm}
{\large Shin Muroya
{\footnote {E-mail address:muroya@yukawa.kyoto-u.ac.jp}},
Hiroki Nakamura$^{1)}$
{\footnote {E-mail address:64l088@cfi.waseda.ac.jp}},
and  Mikio Namiki$^{1)}$
{\footnote {E-mail address:namiki@cfi.waseda.ac.jp}},
}\\
\vspace{.5cm}
{\it Tokuyama Women's College, \\
Kume, Tokuyama, Yamaguchi 745, Japan}\\

\vspace{.5cm}

{\it $^{1)}$Department of Physics, Waseda University\\
Okubo 3-4-1, Shinjuku-ku, Tokyo 169-50, Japan}\\
\vspace{.5cm}
\end{center}
\vspace{1cm}
\begin{abstract}
We numerically analyze  recent high energy heavy-ion
collision experiments based on a hydrodynamical model with
phase transition and discuss a systematic change of initial
state of QGP-fluid depending on colliding-nuclei's mass.

In a previous paper, we formulated a (3+1)-dimensional
hydrodynamical model for quark-gluon plasma with phase
transition and discussed numerically the space-time evolution
in detail.  We here compare the numerical solution with the
hadronic distributions given by CERN WA80 and NA35.
Systematic analyses of the experiments with various colliding
nuclei enable us to discuss the dependences of the initial
parameters of the hydrodynamical model on colliding nuclei's
mass.
Furthermore, extrapolating the present experiments, we derive
the possible hadronic distributions for lead-lead 150GeV/A collision.

\end{abstract}
\vspace{1cm}

\vspace{2cm}

\baselineskip 24pt
\parskip 12pt
\section{Introduction}

Physics of quark-gluon plasma is one of the most important problems
in high energy physics.
Quark-gluon plasma state is expected to be produced in extremely
high temperature, which will be realized in ultra-relativistic
heavy ion collisions.
To obtain the higher temperature, experimental setting has been
growing in three directions: the higher incident energy par nucleon,
the heavier incident nucleus, and the heavier target nucleus.
In any case, colliding energy is getting larger, but the larger energy
does not directly mean the higher temperature.  It is not obvious,
whether large fraction of colliding energy will be transferred into thermal
energy (temperature) or kinetic energy (systematic flow).  Especially,
the heavier target does not correspond to the larger colliding velocity
in the center of mass system.  Hence, if we stand on the Landau-type
``simple stopping" picture, achieved energy density must become lower
for larger target mass.
In this paper, we will analyze the recent experimental data for various
target nuclei and discuss systematic change of initial parameters
in hydrodynamical model.

In a previous paper\cite{mmn} we formulated a semi-phenomenological
quantum transport theory for a quark-gluon plasma fluid based on an
operator valued Langevin equation.  Putting a mode spectrum and a
damping into this formula, we can easily evaluate thermodynamical
quantities and transport coefficients.
Introducing a simple model spectrum to this formula,  we have already
discussed the space-time evolution of (1+1)-dimensional viscous
quark-gluon plasma fluid \cite{date} and (3+1)-dimensional perfect fluid
quark-gluon plasma with phase-transition \cite{akase}.   We have also
discussed the case of baryon-rich quark-gluon plasma with
phase-transition \cite{ishii}, in which one of the conclusions was that
the baryon-number effect was not so much large. Hence, although the
existing experiments do not seem to be baryon-free, we apply a simple
baryon-free quark-gluon plasma model to analyses of available experiments.

\section{Hydrodynamical model with phase-transition}

Following \cite{date}, we here introduce a simple model mode spectrum,
\begin{equation}
\varepsilon({\bf k})=A\sqrt{{\bf k}^{2}+M^{2}}
\frac{1-\tanh\frac{T-T_{\rm C}}{d}}{2}
+|{\bf k}|\frac{1+\tanh\frac{T-T_{\rm C}}{d}}{2}
\end{equation}
where $A$ is a parameter that adjusts the degrees of modes and $T_{\rm C}$
is the critical temperature responsible for the reliant phase transition.
Supposing that the high temperature phase is dominated by massless u-, d- ,
s-quarks and gluons and the hadronic phase is dominated by pions and kaons,
we can put $ A = 1.89 $, $M= 200$ MeV, $d= 2$ MeV, and $T_{\rm C}=160$ MeV.
With these parameters, we can easily obtain an equation of state with phase
transition-like behavior (fig.1), which seems to reproduce the Lattice QCD
result \cite{akase}.

The hydrodynamical equation is given by,
\begin{eqnarray}
\partial_{\mu}T^{\mu\nu}&=&0,
\label{1}
\\
T^{\mu\nu}&=& EU^{\mu}U^{\nu}-P(g^{\mu\nu}-U^{\mu}U^{\nu}),
\end{eqnarray}
\noindent
for perfect fluid.  Here $E$,$P$
and $U^{\mu}$ are, respectively, energy density, pressure and local
four velocity.
For cylindrically symmetric expansion along the collision axis, it is
convenient to introduce new variables, $\tau$,\ $\eta$,\ $r$ and $\phi$
defined by
\begin{eqnarray}
t&=&\tau\cosh \eta, \nonumber
\\
z&=&\tau\sinh \eta, \nonumber
\\
x&=&r \cos \phi, \nonumber
\\
y&=&r \sin \phi, \nonumber
\end{eqnarray}
\noindent
instead of ordinary coordinate $t$,\ $x$,\ $y$ and $z$.  As for the local
four velocity, we represent it by four components, $U^{\tau}$,\ $U^{\eta}$,
\
$U^{r}$ and $U^{\phi}$ associated with the new variables,
\begin{eqnarray}
U^{t}&=&U^{\tau}\cosh \eta +U^{\eta}\sinh \eta, \nonumber
\\
U^{z}&=&U^{\tau}\sinh \eta +U^{\eta}\cosh \eta, \nonumber
\\
U^{x}&=&U^{r}\cos \phi+U^{\phi}\sin \phi, \nonumber
\\
U^{y}&=&U^{r}\sin \phi-U^{\phi}\cos \phi. \nonumber
\end{eqnarray}
\noindent
Because of $U^{\mu}U_{\mu}=1$ and the cylindrical symmetry, the four components
can be reduced to two variables $Y_T$ and $Y_L$\cite{akase},
\begin{eqnarray}
U^{\tau}&=&\cosh Y_{\rm T} \cosh(Y_{\rm L}-\eta), \nonumber
\\
U^{\eta}&=&\cosh Y_{\rm T} \sinh(Y_{\rm L}-\eta), \nonumber
\\
U^{r}&=&\sinh Y_{\rm T}, \nonumber
\\
U^{\phi}&=&0 . \nonumber
\end{eqnarray}

Taking account of the formation time, let us put
the temperature distribution on $\eta$ and $r$ at $\tau=\tau_0$,
\begin{equation}
T(\tau_0,\eta)=T_0\exp\bigl(-{\frac{(\vert \eta \vert-\eta_0)^2}
{3 \cdot2 \cdot{\sigma_\eta}^2}} \theta(\vert \eta \vert-\eta_0)
-{\frac{(r-r_0)^2}{3 \cdot2 \cdot{\sigma_r}^2}} \theta(r-r_0)\bigr)
\label{(3)},
\end{equation}
assuming that the hydrodynamical expansion starts at $\tau=\tau_0 = 1$ fm later
than the collision instance at $\tau=0$ fm.  $T_0$ represents the initial
temperature,
$\eta_0$ and $r_0$ are measures of the longitudinal and transverse spreads
in $\eta$ and $r$, respectively.
$T_0$, $\eta_0$, $r_0$,$\sigma_{\eta}$ and $\sigma_r$
are input parameters to characterize our model on which we should
impose a constraint given by a fixed value of the initial fluid energy
$E_{ini}=\chi E_{tot}$ ($E_{tot}$ standing for the total collision energy and
$\chi$ for the inelasticity).  As for the initial condition of the local
velocity, we use Bjorken's scaling solution, $Y_L = \eta$, in the
longitudinal direction, by which we can take the initial longitudinal flow
into account.
Initial values of these parameters should be determined on the basis of
an appropriate physical discussion, but details of these parameters are deeply
connected to the interaction mechanism and the thermalization
process.
In this paper, standing on {\it phenomenological} point of view, we
try to choose these parameters so as to reproduce experimental results of both
momentum spectrums.

\noindent
\section{Particle Distributions}

The numerical solution of the hydrodynamical equation gives us
the momentum
distribution of hadrons, coming out from a local system with volume
$d \sigma^{\mu}$ in local equilibrium with freeze-out temperature
$T_f$,
through the formula
\begin{equation}
\frac{d^3\Delta N}{d {\bf p}^3}
=\frac{U^{\mu}p_{\mu}}{\sqrt{ {\bf p}^2 +m^2}}
\frac{1}{\exp {\bigl( \frac{U^{\rho}p_{\rho}}{T_{\rm f}} \bigr)}-1}
\frac{U^{\nu}d\sigma_{\nu}}{(2 \pi)^3}
\label{5.3}
\end{equation}
where $p^0 = \sqrt{p^2+m^2}$ with hadron mass $m$ (see ref.\cite{akase}).
Integrating eq.(\ref{5.3}) on the hypersurface with $T_f$,
we obtain
\begin{equation}
\frac{dN}{dY}
=\int \frac{U^{\mu}p_{\mu}}
{\exp {\bigl(\frac{U^{\rho}p_{\rho}}{T_{\rm f}}\bigr)-1}}
\frac{U^{\nu}d\sigma_{\nu}P_{\rm T}dP_{\rm T}d\varphi}{(2 \pi)^3}
\label{5.4}
\end{equation}
for the Y-distribution,
\begin{equation}
\frac{dN}{ d{\eta}'}
=\int \frac{1}{\sqrt{ {\bf p}^2 +m^2}}
\frac{1}{\exp {\bigl(\frac{U^{\mu}p_{\mu}}{T_{\rm f}}\bigr)}-1}
\frac{U^{\nu}d\sigma_{\nu}dP_{\rm L}d\varphi}{(2 \pi)^3},
\end{equation}
for ${\eta}'$-distribution,
where ${\eta}' = \frac{1}{2}{\rm ln}\frac{P-P_{L}}{P+P_{L}}$ stands for the
pseudorapidity distribution, and
\begin{equation}
\frac{dN}{P_{\rm T}dP_{\rm T}}
=\int \frac{U^{\mu}p_{\mu}}{\sqrt{ {\bf p}^2 +m^2}}
\frac{1}{\exp {\bigl(\frac{U^{\mu}p_{\mu}}{T_{\rm f}}\bigr)}-1}
\frac{U^{\nu}d\sigma_{\nu}dP_{\rm L}d\varphi}{(2 \pi)^3}
\label{5.5}
\end{equation}
for the $P_T$-distribution.

Our numerical results given by eq.(\ref{5.5}) is shown in fig.2.
As is well known, the freeze-out temperature $ T_f$ is directly
connected with the $P_T$ slope and we should determine it from the
experimental $P_T$-spectrum.  However, throughout this paper, we fix it
equal to $ T_f=140$MeV  because of the lack of sufficient experimental data.

Our numerical solution of hydrodynamical model is so designed as to describe
the head-on collision of the same kind of nuclei in the center of mass
system.
In order to apply the solution to the asymmetric collision, such as WA80
and NA35, we estimated effective participants in large target nucleus as,
\begin{equation}
A_{eff}=A_{tar} \left[ 1-\left(1-\left(\frac{A_{proj}}{A_{tar}}
\right)^{\frac{2}{3}} \right)^{\frac{3}{2}} \right],
\end{equation}
$A_{tar}$ and $A_{proj}$ being target mass and projectile mass,
respectively.

\section{Numerical results}

Our hydrodynamical model contains many parameters to be adjusted.  We have
already fixed freeze-out temperature as $ T_f=140$MeV from fig.2 in
the previous section. We may put formation time equal to $\tau_{0} = 1$ fm
and initial transverse size equal to the size of smaller nucleus,

\begin{equation}
r_{0} = 1.2\times \sqrt[3]{A_{proj}} - \sigma_{r}  \mbox{\quad (fm)},
\end{equation}
where $ \sigma_{r} = 1 \mbox{\quad (fm)}$.
As for the local velocity, we use Bjorken's scaling solution,
$Y_L = \eta$, neglecting initial transverse flow.
Therefore, the residual parameters to be adjusted are only initial
temperature $T_{0}$ at
central point , and longitudinal size at initial time, $\eta_{0}$
and $\sigma_{\eta}$. We try to adjust these parameters
so as to reproduce existing experiments.  Our results for sulphur beam
experiments are shown in fig.3, and the values of parameters for sulphur
and
for oxygen beam are summarized in table 1.
In any case, our model seems to reproduce the experimental results well
with plausible values of the parameters. Table 2 stands for the
results given by
a equation of state in the limit $T_{C} \to \infty$, which corresponds to a
hot hadron-gas model.  In this case, $\chi$ became an unphysical value, larger
than unity.  It means that {\it we failed to reproduce experimental
results with the hot hadron-gas model}.

These results tell us that larger nuclei are more effective to make
a high temperature fluid than smaller nuclei can do even in target experiment.
For larger targets, the initial temperature must become higher,
the longitudinal
size smaller, and the inelasticity larger.   This kind of tendency
is just desired for the experimental quark-gluon plasma production.

Sometimes, it has been pointed out that the present CERN energy is not
high enough to make the Bjorken's scaling quark-gluon plasma, and
the stopping picture of Landau type is then realized.
However, the simple stopping
picture is not in agreement with our results.  In the simple stopping picture,
the only collision parameter is the colliding relative velocity $\gamma$,
 and the longitudinal size contracts with $1/\gamma$ and the energy density
increases with  ${\gamma}^2$.    In target experiments, the larger target
mass means the higher collision energy, but does not means the higher
collision velocity.   Table 3 shows the our results and relative
velocities.  The tendency is opposite
to the simple Landau
picture.   Of course,  it only means the failure of the {\it simple}
stopping picture and we may describe the phenomena which require to
introduce additional
phenomenological parameters, such as energy dependent inelasticity
\cite{weiner}.

In this paper, we have fitted hydrodynamical parameters in a purely
phenomenological manner without resorting to any
dynamical model for initial collision process.  Although
we need a dynamical model for the quark-gluon plasma formation to understand
the results in the table 1, we may extrapolate our phenomenological
results  in order to obtain the rough sketch of coming experiments.
Figure 4 indicates us that parameters in Pb + Pb 150 GeV/A collision are
$T_{0} = 250$MeV and $\eta_{0}+\sigma_{\eta} = 0.8$.  The possible
psuedorapidity-distribution of hadrons is shown in fig. 5 and the total
charged hadronic multiplicity  (pions and kaons) reaches 2095.

\section{Concluding Remarks}

We have applied our (3+1)-dimensional hydrodynamical model
with phase-transition
to the recent heavy-ion experiments.  Our results show that the inelasticity
increases with target mass, the longitudinal size is getting smaller with
target mass, and the initial temperature becomes higher.
Such a tendency is very promising for our aim of producing quark-gluon plasma
experimentally and also for understanding the future experimental results
at CERN and RHIC.

The authors are indebted to Professor I. Ohba for his suggestive discussions.

\newpage
\begin{small}

\end{small}

\newpage
\begin{center}
\begin{tabular}{c|c|c|c|c|c}
\multicolumn{6}{c}{\bf Table 1. Phase-Transition Model }\\
\hline
\hline
Collision  & \quad $ A $ \quad & \quad $ \sigma_{\eta} $ \quad &
\quad $ \eta_{0} $ \quad & $T_{0}$(MeV)& $ \chi $
\\ \hline
S + Al ${^{1)}}$ & 27$^{\dagger}$  & 1.5 & 1.0 & 169 & 0.52 \\ \hline
S + S  $\ {^{2)}}$  & 32  & 1.5 & 1.0 & 167 & 0.46 \\ \hline
S + Cu ${^{1)}}$ & 64  & 1.5 & 0.8 & 180 & 0.69 \\ \hline
S + Ag ${^{1)}}$ & 107 & 1.5 & 0.7 & 188 & 0.75 \\ \hline
S + W  ${^{3)}}$  & 184 & 1.2 & 0.5 & 202 & 0.68 \\ \hline
S + Au ${^{1)}}$ & 197 & 1.5 & 0.7 & 195 & 0.88 \\ \hline
O + Cu ${^{4)}}$& 64 & 167  & 3.0 & 0.3 & 0.38 \\ \hline
O + Ag ${^{4)}}$& 107 & 172  & 3.0 & 0.2 & 0.54 \\ \hline
O + Au ${^{4)}}$& 197 & 176  & 3.0 & 0.1 & 0.62\\ \hline
\multicolumn{6}{c}{
Data 1) from \cite{WA80}, 2) from \cite{NA35}, 3) from \cite{HELIOS} and 4)
from \cite{Oxygen}
 } \\
\end{tabular}
\end{center}

\begin{center}
\begin{tabular}{c|c|c|c|c|c}
\multicolumn{6}{c}{\bf Table 2.
Hadron Model}\\
\hline
\hline
Target & \quad $ A $ \quad & \quad $ \sigma_{\eta} $ \quad &
\quad $ \eta_{0} $ \quad & $T_{0}$(MeV)& $ \chi $
\\ \hline
Al${^{1)}}$ & 27$^{\dagger}$  & 0.8 & 0.5 & 335 & 0.82 \\ \hline
S$\ {^{2)}}$  & 32  & 0.6 & 1.0 & 312 & 0.93 \\ \hline
Cu${^{1)}}$ & 64  & 1.0 & 0.5 & 363 & 1.37 \\ \hline
Ag${^{1)}}$ & 107 & 1.0 & 0.5 & 380 & 1.49 \\ \hline
W$\ {^{3)}}$  & 184 & 0.8 & 0.7 & 405 & 1.65 \\ \hline
Au${^{1)}}$ & 197 & 0.9 & 0.7 & 400 & 1.77 \\ \hline
\multicolumn{6}{c}{
Data 1) from \cite{WA80}, 2) from \cite{NA35} and 3) from \cite{HELIOS}} \\
\end{tabular}
\end{center}

\begin{center}
\begin{tabular}{c|c|c|c|c|c}
\multicolumn{6}{c}{\bf Table 3. Parameters for Landau Picture }\\
\hline
\hline
Collision & Effective & $E_{lab}$ & $E_{CM}$ & $ \gamma_{cm}$ & $\sigma_{\eta}
+\eta_{0}$ \\
 &  Participants &  (GeV) &    (GeV)&  &  \\ \hline
S + Al & 30.9 + 27 & 6175.72 & 561.218 & 11 & 1.5 + 1.0 \\ \hline
S + Cu & 32 + 49.6 & 6400 & 774.461 & 8.3 & 1.5 + 0.8 \\ \hline
S + Ag & 32 + 63.0 & 6400 & 873.332 & 7.4 & 1.5 + 0.7 \\ \hline
S + Au & 32 + 81.1 & 6400 & 990.954 & 6.5 & 1.5 + 0.7 \\ \hline \hline
O + Cu & 16 + 34.0 & 3200 & 453.783 & 7.1 & 3.0 + 0.3 \\ \hline
O + Ag & 16 + 41.9 & 3200 & 503.616 & 6.4 & 3.0 + 0.2 \\ \hline
O + Au & 16 + 52.7 & 3200 & 565.62  & 5.7 & 3.0 + 0.1 \\ \hline
\end{tabular}
\end{center}

\newpage
\noindent
{\large \bf Figure Caption}

\noindent
{\bf Fig.1}

\noindent
The energy density as a function of temperature; the solid line stands for our
phase-transition model, the dotted line for the simple hadron-gas of pions and
kaons, and the dot-dashed line for the quark-gluon plasma of massless u-, d-,
s-quarks and gluons.
In the phase-transition model, phase-transition takes place at $T_{C}=160$MeV
in the width of $d=2$MeV.

\vspace{.1in}
\noindent
{\bf Fig.2}

\noindent
The $P_{T}$ spectrum of pions in S+S 200 GeV/A collision together with the data
of NA35. The solid line stands for $T_{f}=140$ MeV, the dashed line for
$T_{f}=120$MeV, the dotted line for $T_{f}=100$MeV, and dot-dashed line for
$T_{f}=80$MeV, respectively.

\vspace{.1in}
\noindent
{\bf Fig.3}

\noindent
The pseudorapidity distributions of charged hadron in 200GeV/A Sulphur beam
collision. The solid line stands for S+Al, the dashed line for S+Cu, the dotted
line for S+Ag, and the dot-dashed line stands for S+Au, respectively. In our
calculation, we take account of pions and kaons only.  Plots are the data from
\cite{WA80}. In the case of S+Al, $r_{0}$ is determined from the size of Al.

\vspace{.1in}
\noindent
{\bf Fig.4}

\noindent

The values of initial parameters as a function of Total Collision Energy.
The solid line stands  for initial temperature,$T_{0}$, and the dashed line for
longitudinal extent, $\eta_{0}+\sigma_{\eta}$.
The plots from the left to the right correspond to O+Cu, O+Ag, O+Au, S+Al,
S+Cu, S+Ag, and S+Au, respectively.  Plots at the right end are the
extrapolation of parameters to Pb+Pb 150 GeV/A collision.

\vspace{.1in}
\noindent
{\bf Fig.5}

\noindent
The psuedorapidity distribution of charged hadrons ($\pi^{\pm}$ and $K^{\pm}$ )
in Pb+Pb 150 GeV/A collision.  Here, we put $T_{0}=$250MeV,  $\eta_{0}=$0.2,
and $\sigma_{\eta}= 0.6$.

\end{document}